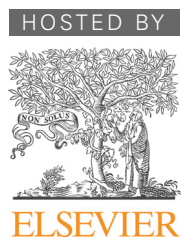

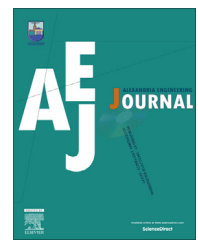

# ARA-residual power series method for solving partial fractional differential equations

**Aliaa Burqan** [a,*], **Rania Saadeh** [a], **Ahmad Qazza** [a], **Shaher Momani** [b,c]

[a] *Department of Mathematics, Zarqa University, Zarqa 13110, Jordan*
[b] *Nonlinear Dynamics Research Center (NDRC), Ajman University, Ajman, United Arab Emirates*
[c] *Department of Mathematics, The University of Jordan, Amman 11942, Jordan*

**Abstract** In this article a new approach in solving time fractional partial differential equations (TFPDEs) is introduced, that is, the ARA-residual power series method. The main idea of this technique, depends on applying the ARA-transform and using Taylor's expansion to construct approximate series solutions. The procedure of getting the approximate solutions for nonlinear TFPDEs is a difficult mission, the ARA-residual power series method over comes this trouble throughout expressing the solution in a series form then obtain the series coefficients using the idea of the residual function and the concept of the limit at infinity. This method is efficient and applicable to solve a wide family of TFPDEs. Four attractive applications are considered to show the speed and the strength of the proposed method in constructing solitary series solutions of the target equations.

## 1. Introduction

Numerous phenomena in various fields of science can be fruitfully formulated by using fractional derivatives. That is because sensible modeling for physical phenomena depends on instantaneous time as well as on prior time history. Hence, Many physical and engineering problems can be formulated by fractional differential equations (FDEs) and obtaining the solutions of these equations have been the theme of many interesting investigations [1–19].

The power series method (PSM) is used to solve some classes of linear fractional differential equations. It is based on assuming that the solution of equation can be expressed as a power series that leads to a closed solution form of the exact solution. This method has received attention of many researchers and it is implemented to solve different types of linear fractional differential equations [20–25]. Unfortunately, the procedure of getting the approximate solutions for nonlinear fractional differential equations and the determining of the series coefficients is a very difficult mission. To overcome the mentioned troubles, an improvement of the PSM is introduced; the residual power series method (RPSM) is established to treat the coefficients as transformed functions by adopting the derivatives in determining it [25]. Residual power series method has been used to solve analytically many important models of linear and nonlinear equations arising in different disciplines of engineering and science fields. In other development, the Laplace residual power series method (LRPSM) is established in 2020 [26,27] by combining the Laplace transform with RPSM. In this article, another promotion of the RPSM is constructed by adapting the ARA transform [28,29] in the methodology of RPSM [30,31]. This powerful technique, is a new scheme in finding approximate solutions of nonlinear partial differential equations of fractional order in a series form. The series coefficients can be determined rapidly using the concept of limit at infinity, which reduces the time and efforts doing the calculations in a comparison with other methods. The proposed method, ARA-residual power series method (ARA-RPSM) is implemented in this article to solve nonlinear TFPDEs of the form:

$$D_t^\alpha y(x,t) = \mathcal{N}_x[y(x,t)], \ 0 < \alpha \le 1, \quad x \in I \ \text{ and } \ t > 0 \quad (1)$$

with initial conditions (ICs)

$$y(x,0) = a(x), \ \ D_t^\alpha y(x,0) = b(x). \quad (2)$$

where $N_x$ is a nonlinear operator relative to $x$ of degree $r$, $D_t^\alpha$ is the Caputo differential operator of order $\alpha$, and $y(x,t)$ is a known function of $x$ and $t$.

* Corresponding author.
E-mail address: aliaaburqan@zu.edu.jo (A. Burqan).
Peer review under responsibility of Faculty of Engineering, Alexandria University.

On the other hand, it is prominence mention that there are many operative analytical and numerical methods have contributed appreciably in scientific applications. Some of these methods include, for example: Adomian's decomposition method, homotopy perturbation and analysis methods, different forms of the fractional power series representation, differential transform method and its modifications, iterative methods and among others [32–64].

This article is organized as follows. In Section 2, we review some essential concepts, definitions and results related to the fractional derivatives and the ARA transform. In Section 3, the ARA-RPSM is implemented to construct and predict solutions of nonlinear TFPDEs. In Section 4, the capability, simplicity and efficiency of the proposed method are tested by solving examples included four different types of TFPDEs. Finally in the conclusion section, our results are summarized.

## 2. Basic definitions and theorems

In this section. The definitions of the fractional operators Riemann–Liouville and Caputo are introduced. Also, the definition of the ARA transform, and some properties and theorems related to the fractional ARA-RPSM are revisited.

**Definition 1.** *The Caputo fractional derivative of the function* $y(x,t)$ *of order* $\alpha > 0$ *is defined by*

$$D_t^\alpha y(x,t) = J_t^{m-\alpha} D_t^m y(x,t), \ m-1 < \alpha < m, \ m \in n, \ x \in I, \ t > 0,$$

*where* $I$ *is an interval and* $J_t^\alpha$ *is the time- fractional Riemann–Liouville integral operator of order* $\alpha > 0$, *defined as.*

$$J_t^\alpha y(x,t) = \begin{cases} \frac{1}{\Gamma(\alpha)} \int_0^t (t-\tau)^{\alpha-1} y(x,\tau)d\tau, & t > \tau > 0 \\ y(x,\tau), & \alpha = 0, \end{cases}$$

**Definition 2** *([28]). The ARA transform of order n of the continuous function* $y(x,t)$ *on the interval* $I \times [0,\infty)$ *for the veriable* $t$, *is defined by*

$$\mathcal{G}_n[y(x,t)] = s \int_0^\infty t^{n-1} e^{-st} y(x,t)dt, \ s > 0.$$

In the following arguments, we present some basic properties of the ARA transform [28] that are essential in our research.

**Properties of the ARA-transform** [28–30]

Let $y(x,t)$ and $h(x,t)$ be continuous functions on $I \times [0,\infty)$ in which the ARA-transform for the variable $t$ exists. Then we have.

(1) $\mathcal{G}_n[ay(x,t) + bh(x,t)] = a\mathcal{G}_n[y(x,t)] + b\mathcal{G}_n[h(x,t)]$, where $a$ and $b$ are nonzero constants.

(2) $\lim\limits_{s\to\infty} \mathcal{G}_1[y(x,t)] = y(x,0), x \in I, s > 0$.

(3) $\mathcal{G}_1[D_t^\alpha y(x,t)] = s^\alpha \mathcal{G}_1[y(x,t)] - s^\alpha y(x,0), \ 0 < \alpha \le 1, x \in I, s > 0$

(4) $\mathcal{G}_2[t^\alpha] = \frac{\Gamma(\alpha+2)}{s^{\alpha+1}}, \ \alpha > 0, s > 0$.

(5) $\mathcal{G}_2[D_t^\alpha y(x,t)] = s^\alpha \mathcal{G}_2[y(x,t)] - \alpha s^{\alpha-1}\mathcal{G}_1[y(x,t)]$
$\quad + (\alpha-1)s^\alpha y(x,0), \ 0 < \alpha \le 1, \ x \in I, s > 0$.

(6) $\mathcal{G}_2[D_t^{2\alpha}y(x,t)] = s^{2\alpha}\mathcal{G}_2[y(x,t)] - 2\alpha s^{2\alpha-1}\mathcal{G}_1[y(x,t)]$
$\quad + (2\alpha-1)s^{2\alpha-1}y(x,0)$
$\quad + (\alpha-1)s^{\alpha-1}D_t^\alpha y(x,0), 0 < \alpha \le 1, x \in I, s > 0$.

(7) $\lim\limits_{s\to\infty} s\mathcal{G}_2[y(x,t)] = y(x,0), x \in I, s > 0$.

**Theorem 1** *([30]). Suppose that the fractional power series (FPS) representation of the function* $y(x,t)$ *at* $t = 0$ *has the form*

$$y(x,t) = \sum_{n=0}^\infty a_n(x)t^{n\alpha}, \quad m-1 < \alpha \le m, \quad m = 1,2,\cdots, \quad 0 \le t \le \beta.$$

*If* $y(x,t)$ *and* $D_t^{n\alpha}y(x,t)$ *are continuous on* $I \times [0,\infty)$, *then the coefficients* $a_n(x)$ *have the form*

$$a_n(x) = \frac{D_t^{n\alpha}y(x,0)}{\Gamma(n\alpha+1)}, \ \text{ for } n = 0,1,2,\cdots$$

*where* $D_t^{n\alpha} = D_t^\alpha \cdot D_t^\alpha \cdots D_t^\alpha (n- \text{ times})$.

**Theorem 2** *([30]). Let* $y(x,t)$ *be a continuous function on* $I \times [0,\beta]$ *in which the ARA-transform for the variable* $t$ *exists and has the FPS representation*

$$\mathcal{G}_2[y(x,t)] = \sum_{n=0}^\infty \frac{h_n(x)}{s^{n\alpha+1}}; \ 0 < \alpha \le 1, \ x \in I \text{ and } s > 0. \quad (3)$$

*Then*

$$h_n(x) = (n\alpha+1)D_t^{n\alpha}y(x,0). \quad (4)$$

**Remark 1.**

i. *The* $k^{th}$ *truncated series of the series representation* (3) *is defined as follows*

$$\mathscr{G}_2[y(x,t)]_k = \sum_{n=0}^{k} \frac{h_n(x)}{s^{n\alpha+1}}. \tag{5}$$

ii. *If the ARA-transform of order two of the function $y(x,t)$ has the series representation (3), then the ARA-transform of order one can be expressed as follows*

$$\mathscr{G}_1[y(x,t)] = \sum_{n=0}^{\infty} \frac{h_n(x)}{(n\alpha+1)s^{n\alpha}}, \tag{6}$$

*and the $k^{th}$ truncated series is defined as follows:*

$$p\mathscr{G}_1[y(x,t)]_k = \sum_{n=0}^{k} \frac{h_n(x)}{(n\alpha+1)s^{n\alpha}}. \tag{7}$$

iii. *The inverse of the ARA-transform of order two for the fractional power series (3) is*

$$y(x,t) = \mathscr{G}_2^{-1}\left[\sum_{n=0}^{\infty} \frac{h_n(x)}{s^{n\alpha+1}}\right](t) = \sum_{n=0}^{\infty} \frac{D_t^{n\alpha}y(x,0)}{\Gamma(n\alpha+1)} t^{n\alpha}.$$

**Theorem 3** (*[30]*). *Let $y(x,t)$ be a continuous function on $I \times [0,\beta]$ in which the ARA-transform for the variable $t$ exists. Assume that $\mathscr{G}_1[y(x,t)]$ has the following series representation:*

$$\mathscr{G}_1[y(x,t)] = \sum_{n=0}^{\infty} \frac{C_n(x)}{s^{n\alpha}},$$

*where*

$$C_n(x) = D_t^{n\alpha}y(x,0).$$

*If $\left|\mathscr{G}_1\left[D_t^{(n+1)\alpha}y(x,t)\right]\right| \leq M$ on $0 < s \leq d$, then the remainder $R_n(x,s)$ satisfies the following inequality:*

$$|R_n(x,s)| \leq \frac{M(x)}{s^{(n+1)\alpha}}, \ x \in I, \ 0 < s \leq d.$$

## 3. Constructing series solutions of TFPDEs

In this section, we present the methodology of the ARA-RPSM for solving nonlinear PDEs of time fractional type. The main idea of the proposed method is based on applying the ARA-transform on the given equation and using Taylor's expansion to create solitary solutions.

To perform the ARA-RPSM, consider the initial value problem (IVP) (1) and (2).

Operate the ARA-transform of order two $\mathscr{G}_2$ with respect to the variable $t$, on both sides of equation (1).

$$\mathscr{G}_2\left[D_t^{\alpha}y(x,t)\right] = \mathscr{G}_2[\mathscr{N}_x[y(x,t)]]. \tag{8}$$

Using property 6 and the ICs (2), equation (8) becomes.

$$\mathscr{G}_2[y(x,t)] - \frac{2\alpha}{s}\mathscr{G}_1[y(x,t)] + \frac{(2\alpha-1)}{s}a(x) + \frac{(\alpha-1)}{s^{\alpha+1}}b(x) - \frac{1}{s^{2\alpha}}\mathscr{G}_2\left[\mathscr{N}_x\left(\mathscr{G}_2^{-1}[\mathscr{G}_2[y(x,t)]]\right)\right]$$
$$= 0. \tag{9}$$

Assume that the ARA-RPS solution of equation (9) has the following series representations.

$$\mathscr{G}_1[y(x,t)] = \sum_{n=0}^{\infty} \frac{h_n(x)}{(n\alpha+1)s^{n\alpha}}, \tag{10}$$

$$\mathscr{G}_2[y(x,t)] = \sum_{n=0}^{\infty} \frac{h_n(x)}{s^{n\alpha+1}}. \tag{11}$$

Using the fact in property 7.

$$\lim_{s\to\infty} s\mathscr{G}_2[y(x,t)] = y(x,0),$$

we have $h_0(x) = a(x)$. Hence the series representation (11) becomes.

$$\mathscr{G}_2[y(x,t)] = \frac{a(x)}{s} + \frac{h_1(x)}{s^{\alpha+1}} + \sum_{n=2}^{\infty} \frac{h_n(x)}{s^{n\alpha+1}}. \tag{12}$$

To find $h_1(x)$, we multiply both sides of equation (12) by $s^{\alpha+1}$ and take the limit as $s \to \infty$ to obtain.

$$\lim_{s\to\infty} s^{\alpha+1}\mathscr{G}_2[y(x,t)] = \lim_{s\to\infty} s^{\alpha}a(x) + h_1(x) + \lim_{s\to\infty} \sum_{n=2}^{\infty} \frac{h_n(x)}{s^{n\alpha+1}},$$

which is equivalent to.

$$h_1(x) = \lim_{s\to\infty} s\left(s^{\alpha}\mathscr{G}_2[y(x,t)] - s^{\alpha-1}a(x)\right).$$

Property 5 yields that.

$$h_1(x) = \lim_{s\to\infty} s\left(\mathscr{G}_2\left[D_t^{\alpha}y(x,t)\right] + \alpha s^{\alpha-1}\mathscr{G}_1[y(x,t)] - \alpha s^{\alpha-1}a(x)\right)$$
$$= \lim_{s\to\infty} s\mathscr{G}_2\left[D_t^{\alpha}y(x,t)\right] + \lim_{s\to\infty} \alpha(s^{\alpha}\mathscr{G}_1[y(x,t)] - s^{\alpha}a(x)).$$

Using property 3, we get.

$$h_1(x) = \lim_{s\to\infty} s\mathscr{G}_2\left[D_t^{\alpha}y(x,t)\right] + \alpha\lim_{s\to\infty} \mathscr{G}_1\left[D_t^{\alpha}y(x,t)\right].$$

Property 2 and 7 lead to.

$$h_1(x) = (\alpha+1)D_t^{\alpha}b(x).$$

Thus, the ARA-RPS solution of equation (9) has the series representations.

$$\mathscr{G}_1[y(x,t)] = a(x) + \frac{b(x)}{s^{\alpha}} + \sum_{n=2}^{\infty} \frac{h_n(x)}{(n\alpha+1)s^{n\alpha}}, \tag{13}$$

$$\mathscr{G}_2[y(x,t)] = \frac{a(x)}{s} + \frac{(\alpha+1)b(x)}{s^{\alpha+1}} + \sum_{n=2}^{\infty} \frac{h_n(x)}{s^{n\alpha+1}}, \tag{14}$$

and the $k^{th}$ truncated series expansion of equation (13) and equation (14) have the form.

$$\mathscr{G}_1[y(x,t)]_k = a(x) + \frac{b(x)}{s^{\alpha}} + \sum_{n=2}^{k} \frac{h_n(x)}{(n\alpha+1)s^{n\alpha}}, \tag{15}$$

$$\mathscr{G}_2[y(x,t)]_k = \frac{a(x)}{s} + \frac{(\alpha+1)b(x)}{s^{\alpha+1}} + \sum_{n=2}^{k} \frac{h_n(x)}{s^{n\alpha+1}}. \tag{16}$$

To find the coefficients of the series expansions in equations (15) and (16), we define the ARA-residual function of equation (9) as follows.

$$
\begin{aligned}
\mathscr{G}_2 Res(x,s) = {} & \mathscr{G}_2[y(x,t)] - \frac{2\alpha}{s}\mathscr{G}_1[y(x,t)] + \frac{(2\alpha-1)}{s}a(x) \\
& + \frac{(\alpha-1)}{s^{\alpha+1}}b(x) \\
& + \frac{1}{s^{2\alpha}}\mathscr{G}_2\left[\mathscr{N}_x\big(\mathscr{G}_2^{-1}[\mathscr{G}_2[y(x,t)]]\big)\right],
\end{aligned}
\tag{17}
$$

and the $k^{\text{th}}$ ARA-residual function is.

$$
\begin{aligned}
\mathscr{G}_2 Res_k(x,s) = {} & \mathscr{G}_2[y(x,t)]_k - \frac{2\alpha}{s}\mathscr{G}_1[y(x,t)]_k \\
& + \frac{(2\alpha-1)}{s}a(x) + \frac{(\alpha-1)}{s^{\alpha+1}}b(x) \\
& + \frac{1}{s^{2\alpha}}\mathscr{G}_2\left[\mathscr{N}_x\big(\mathscr{G}_2^{-1}[\mathscr{G}_2[y(x,t)]_k]\big)\right], k=2,3,\cdots.
\end{aligned}
\tag{18}
$$

In order to find the coefficients $h_n(x), n \geq 2$ in the series expansion (16), multiply both sides of equation (18) by $s^{k\alpha+1}, k=2,3,\cdots$, and take the limit as $s \to \infty$, then solve the equations.

$$
\lim_{s\to\infty} s^{k\alpha+1}\mathscr{G}_2 Res_k(x,s) = 0, k=2,3,\cdots.
$$

The following facts are needed to obtain the ARA-RPS solution.

- $\mathscr{G}_2 Res(x,s) = 0, x \in I, s > 0.$
- $\lim\limits_{k\to\infty} \mathscr{G}_2 Res_k(x,s) = \mathscr{G}_2 Res(x,s), x \in I, s > 0.$
- $\lim\limits_{s\to\infty} s\mathscr{G}_2 Res(x,s) = 0$ and $\lim\limits_{s\to\infty} s\mathscr{G}_2 Res_k(x,s) = 0, x \in I, s > 0.$
- $\lim\limits_{s\to\infty} s^{k\alpha+1}\mathscr{G}_2 Res(x,s) = \lim\limits_{s\to\infty} s^{k\alpha+1}\mathscr{G}_2 Res_k(x,s) = 0, \ x \in I, \ s > 0.$

The obtained coefficients $h_n(x)$ are substituted in the series solution (11), then operate the inverse ARA transform of order two $\mathscr{G}_2^{-1}$ to get the solution of the IVP (1) and (2) in the original space.

## 4. Numerical examples

In this section, we introduce four interesting examples of nonlinear TFPDEs. These examples demonstrate the applicability and efficiency of the presented method.

**Example 1.** *Consider the time-fractional Klein–Gordon equation*

$$
\begin{aligned}
D_t^{2\alpha}y(x,t) - v(y^2(x,t))_{xx} + w(y^2(x,t))_{xxxx} = 0, \\
x \in \mathbb{R}, \ t > 0, \ 0 < \alpha \leq 1,
\end{aligned}
\tag{19}
$$

*with the ICs*

$$
\begin{aligned}
y(x,0) = \frac{2\lambda^2}{3v}\left(1 - \cosh\left(\sqrt{\tfrac{v}{w}}\tfrac{x}{2}\right)\right), D_t^\alpha y(x,0) = \frac{2\lambda^3}{3\sqrt{vw}}\left(\sinh\left(\sqrt{\tfrac{v}{w}}\tfrac{x}{2}\right)\right), \\
\text{where} \ \ v,w > 0 \ \text{ and } \ \lambda \in \mathbb{R}.
\end{aligned}
\tag{20}
$$

**Solution**

Applying the ARA-transform of order two $\mathscr{G}_2$, on both sides of the equation (19), we get.

$$
\mathscr{G}_2\big[D_t^{2\alpha}y(x,t)\big] - \mathscr{G}_2\big[v(y^2(x,t))_{xx}\big] + \mathscr{G}_2\big[w(y^2(x,t))_{xxxx}\big] = 0.
\tag{21}
$$

Running the ARA-transform $\mathscr{G}_2$ on both sides of equation (21), we obtain.

$$
\begin{aligned}
& s^{2\alpha}\mathscr{G}_2[y(x,t)] - 2\alpha s^{2\alpha-1}\mathscr{G}_1[y(x,t)] + (2\alpha-1)s^{2\alpha-1}y(x,0) \\
& + (\alpha-1)s^{\alpha-1}D_t^\alpha y(x,0) - v\mathscr{G}_2\left[\partial_x^2\big(\mathscr{G}_2^{-1}[\mathscr{G}_2[y(x,t)]]\big)^2\right] \\
& + w\mathscr{G}_2\left[\partial_x^4\big(\mathscr{G}_2^{-1}[\mathscr{G}_2[y(x,t)]]\big)^2\right] \\
& = 0.
\end{aligned}
\tag{22}
$$

Simplifying equation (22), we have.

$$
\begin{aligned}
& \mathscr{G}_2[y(x,t)] - \frac{2\alpha}{s}\mathscr{G}_1[y(x,t)] + \frac{(2\alpha-1)}{s}y(x,0) \\
& + \frac{(\alpha-1)}{s^{\alpha+1}}D_t^\alpha y(x,0) - \frac{v}{s^{2\alpha}}\mathscr{G}_2\left[\partial_x^2\big(\mathscr{G}_2^{-1}[\mathscr{G}_2[y(x,t)]]\big)^2\right] \\
& + \frac{w}{s^{2\alpha}}\mathscr{G}_2\left[\partial_x^4\big(\mathscr{G}_2^{-1}[\mathscr{G}_2[y(x,t)]]\big)^2\right] \\
& = 0.
\end{aligned}
\tag{23}
$$

Assume that the ARA-RPS solution of equation (23) has the following series representations.

$$
\mathscr{G}_1[y(x,t)] = \sum_{n=0}^{\infty} \frac{h_n(x)}{(n\alpha+1)s^{n\alpha}},
\tag{24}
$$

$$
\mathscr{G}_2[y(x,t)] = \sum_{n=0}^{\infty} \frac{h_n(x)}{s^{n\alpha+1}}.
\tag{25}
$$

and the $k^{\text{th}}$ truncated series of the expansions (24) and (25) are.

$$
\mathscr{G}_1[y(x,t)]_k = \sum_{n=0}^{k} \frac{h_n(x)}{(n\alpha+1)s^{n\alpha}},
\tag{26}
$$

$$
\mathscr{G}_2[y(x,t)]_k = \sum_{n=0}^{k} \frac{h_n(x)}{s^{n\alpha+1}}.
\tag{27}
$$

Multiplying both sides of equation (27) by $s$ and taking the limit as $s \to \infty$, we get.

$$
\lim_{s\to\infty} s\mathscr{G}_2[y(x,t)]_k = h_0(x) + \lim_{s\to\infty} \sum_{n=1}^{k} \frac{h_n(x)}{s^{n\alpha}}.
$$

Using the fact.

$$
\lim_{s\to\infty} s\mathscr{G}_2[y(x,t)]_k = y(x,0),
$$

and the ICs in equation (20) we conclude that.

$$
h_0(x) = \frac{2\lambda^2}{3v}\left(1 - \cosh\left(\sqrt{\tfrac{v}{w}}\tfrac{x}{2}\right)\right).
$$

Hance, the series representation (27) becomes.

$$
\begin{aligned}
\mathscr{G}_2[y(x,t)]_k = {} & \frac{2\lambda^2}{3vs}\left(1 - \cosh\left(\sqrt{\tfrac{v}{w}}\tfrac{x}{2}\right)\right) + \frac{h_1(x)}{s^{\alpha+1}} + \sum_{n=2}^{k} \\
& \times \frac{h_n(x)}{s^{n\alpha+1}}.
\end{aligned}
\tag{28}
$$

To find $h_1(x)$, we multiply both sides of equation (28) by $s^{\alpha+1}$ and take the limit as $s \to \infty$, to obtain.

$$\lim_{s\to\infty} s^{\alpha+1}\,\mathscr{G}_2[y(x,t)]_k = \lim_{s\to\infty} s^\alpha \left(\frac{2\lambda^2}{3v}\left(1-\cosh\left(\sqrt{\frac{v}{w}}\frac{x}{2}\right)\right)\right)$$
$$+ h_1(x) + \lim_{s\to\infty}\sum_{n=2}^{k}\frac{h_n(x)}{s^{(n-1)\alpha}}$$

$$\lim_{s\to\infty} s^{\alpha+1}\,\mathscr{G}_2[y(x,t)]_k = \lim_{s\to\infty} s^\alpha \left(\frac{2\lambda^2}{3v}\left(1-\cosh\left(\sqrt{\frac{v}{w}}\frac{x}{2}\right)\right)\right)$$
$$+ h_1(x).$$

Thus,

$$h_1(x) = \lim_{s\to\infty} s\left[s^\alpha\,\mathscr{G}_2[y(x,t)]_k - s^{\alpha-1}\left(\frac{2\lambda^2}{3v}\left(1-\cosh\left(\sqrt{\frac{v}{w}}\frac{x}{2}\right)\right)\right)\right].$$

Property 5, yields that.

$$h_1(x) = \lim_{s\to\infty} s\Big[\mathscr{G}_2\big[D_t^\alpha y(x,t)\big] + \alpha s^{\alpha-1}\mathscr{G}_1[y(x,t)]$$
$$-\alpha s^{\alpha-1}\frac{2\lambda^2}{3v}\left(1-\cosh\left(\sqrt{\frac{v}{w}}\frac{x}{2}\right)\right)\Big]$$
$$= \lim_{s\to\infty} s\,\mathscr{G}_2\big[D_t^\alpha y(x,t)\big]$$
$$+ \lim_{s\to\infty}\alpha\left[s^\alpha\mathscr{G}_1[y(x,t)] - s^\alpha\left(\frac{2\lambda^2}{3v}\left(1-\cosh\left(\sqrt{\frac{v}{w}}\frac{x}{2}\right)\right)\right)\right].$$

Using property 3 we get.

$$h_1(x) = \lim_{s\to\infty} s\,\mathscr{G}_2\big[D_t^\alpha y(x,t)\big] + \alpha\lim_{s\to\infty}\mathscr{G}_1\big[D_t^\alpha y(x,t)\big].$$

Property 2 and 7 lead to.

$$h_1(x) = \frac{(\alpha+1)\lambda^3}{3\sqrt{vw}}\left(\sinh\left(\sqrt{\frac{v}{w}}\frac{x}{2}\right)\right).$$

Thus, the ARA-RPS solution of equation (23) has the following series representations.

$$\mathscr{G}_1[y(x,t)] = \frac{2\lambda^2}{3v}\left(1-\cosh\left(\sqrt{\frac{v}{w}}\frac{x}{2}\right)\right)$$
$$+ \frac{\lambda^3}{3\sqrt{vw}s^\alpha}\left(\sinh\left(\sqrt{\frac{v}{w}}\frac{x}{2}\right)\right) + \sum_{n=2}^{\infty}\frac{h_n(x)}{(n\alpha+1)s^{n\alpha}}, \quad (29)$$

and

$$\mathscr{G}_2[y(x,t)] = \frac{2\lambda^2}{3vs}\left(1-\cosh\left(\sqrt{\frac{v}{w}}\frac{x}{2}\right)\right)$$
$$+ \frac{\lambda^3(\alpha+1)}{3\sqrt{vw}s^{\alpha+1}}\left(\sinh\left(\sqrt{\frac{v}{w}}\frac{x}{2}\right)\right) + \sum_{n=2}^{\infty}\frac{h_n(x)}{s^{n\alpha+1}}, \quad (30)$$

and the $k^{\text{th}}$ truncated series of the expansions (29) and (30) have the form.

$$\mathscr{G}_1[y(x,t)]_k = \frac{2\lambda^2}{3v}\left(1-\cosh\left(\sqrt{\frac{v}{w}}\frac{x}{2}\right)\right)$$
$$+ \frac{\lambda^3}{3\sqrt{vw}s^\alpha}\left(\sinh\left(\sqrt{\frac{v}{w}}\frac{x}{2}\right)\right) + \sum_{n=2}^{k}\frac{h_n(x)}{(n\alpha+1)s^{n\alpha}}, \quad (31)$$

and

$$\mathscr{G}_2[y(x,t)]_k = \frac{2\lambda^2}{3vs}\left(1-\cosh\left(\sqrt{\frac{v}{w}}\frac{x}{2}\right)\right)$$
$$+ \frac{\lambda^3(\alpha+1)}{3\sqrt{vw}s^{\alpha+1}}\left(\sinh\left(\sqrt{\frac{v}{w}}\frac{x}{2}\right)\right) + \sum_{n=2}^{k}\frac{h_n(x)}{s^{n\alpha+1}}, \quad (32)$$

Now, we define the ARA-residual function of equation (23).

$$\mathscr{G}_2Res(x,s) = \mathscr{G}_2[y(x,t)] - \frac{2\alpha}{s}\mathscr{G}_1[y(x,t)]$$
$$+ \frac{(2\alpha-1)}{s}\frac{2\lambda^2}{3v}\left(1-\cosh\left(\sqrt{\frac{v}{w}}\frac{x}{2}\right)\right)$$
$$+ \frac{(\alpha-1)}{s^{\alpha+1}}\frac{\lambda^3}{3\sqrt{vw}}\left(\sinh\left(\sqrt{\frac{v}{w}}\frac{x}{2}\right)\right)$$
$$- \frac{v}{s^{2\alpha}}\mathscr{G}_2\left[\partial_x^2(\mathscr{G}_2^{-1}[\mathscr{G}_2[y(x,t)]])\right]^2$$
$$+ \frac{w}{s^{2\alpha}}\mathscr{G}_2\left[\partial_x^4(\mathscr{G}_2^{-1}[\mathscr{G}_2[y(x,t)]])\right]^2, \quad (33)$$

and the $k^{\text{th}}$ ARA-residual function of equation (33) is.

$$\mathscr{G}_2Res_k(x,s) = \mathscr{G}_2[y(x,t)]_k - \frac{2\alpha}{s}\mathscr{G}_1[y(x,t)]_k + \frac{(2\alpha-1)}{s}$$
$$\times \frac{2\lambda^2}{3v}\left(1-\cosh\left(\sqrt{\frac{v}{w}}\frac{x}{2}\right)\right) + \frac{(\alpha-1)}{s^{\alpha+1}}$$
$$\times \frac{\lambda^3}{3\sqrt{vw}}\left(\sinh\left(\sqrt{\frac{v}{w}}\frac{x}{2}\right)\right)$$
$$- \frac{v}{s^{2\alpha}}\mathscr{G}_2\left[\partial_x^2(\mathscr{G}_2^{-1}[\mathscr{G}_2[y(x,t)]_k])^2\right]$$
$$+ \frac{w}{s^{2\alpha}}\mathscr{G}_2\left[\partial_x^4(\mathscr{G}_2^{-1}[\mathscr{G}_2[y(x,t)]_k])^2\right]. \quad (34)$$

We using the facts.

$$\mathscr{G}_2Res(x,s) = 0,\ \lim_{s\to\infty}\mathscr{G}_2Res_k(x,s) = 0,$$

$$\lim_{s\to\infty}s^{k\alpha+1}\mathscr{G}_2Res(x,s) = \lim_{s\to\infty}s^{k\alpha+1}\mathscr{G}_2Res_k(x,s) = 0,\ k = 2,3,\cdots$$

to find the second unknown coefficient $h_2(x)$ by substituting the second truncated series $\mathscr{G}_1[y(x,t)]_2$ and $\mathscr{G}_2[y(x,t)]_2$ into the second ARA-residual function $\mathscr{G}_2Res_2(s)$ to obtain.

$$\mathscr{G}_2Res_2(s) = \mathscr{G}_2[y(x,t)]_2 - \frac{2\alpha}{s}\mathscr{G}_1[y(x,t)]_2$$
$$+ \frac{(2\alpha-1)}{s}\frac{2\lambda^2}{3v}\left(1-\cosh\left(\sqrt{\frac{v}{w}}\frac{x}{2}\right)\right)$$
$$+ \frac{(\alpha-1)}{s^{\alpha+1}}\frac{\lambda^3}{3\sqrt{vw}}\left(\sinh\left(\sqrt{\frac{v}{w}}\frac{x}{2}\right)\right)$$
$$- \frac{v}{s^{2\alpha}}\mathscr{G}_2\left[\partial_x^2(\mathscr{G}_2^{-1}[\mathscr{G}_2[y(x,t)]_2])^2\right]$$
$$+ \frac{w}{s^{2\alpha}}\mathscr{G}_2\left[\partial_x^4(\mathscr{G}_2^{-1}[\mathscr{G}_2[y(x,t)]_2])^2\right] = 0. \quad (35)$$

Putting

$$\mathscr{G}_1[y(x,t)]_2 = \frac{2\lambda^2}{3v}\left(1-\cosh\left(\sqrt{\frac{v}{w}}\frac{x}{2}\right)\right)$$
$$+ \frac{\lambda^3}{3\sqrt{vw}s^\alpha}\left(\sinh\left(\sqrt{\frac{v}{w}}\frac{x}{2}\right)\right) + \frac{h_2(x)}{(2\alpha+1)s^{2\alpha}},$$

and

$$\mathscr{G}_2[y(x,t)]_2 = \frac{2\lambda^2}{3vs}\left(1 - \cosh\left(\sqrt{\frac{v}{w}}\frac{x}{2}\right)\right)$$
$$+ \frac{\lambda^3(\alpha+1)}{3\sqrt{vw}s^{\alpha+1}}\left(\sinh\left(\sqrt{\frac{v}{w}}\frac{x}{2}\right)\right) + \frac{h_2(x)}{s^{2\alpha+1}},$$

in equation (35) and simplifying, we get.

$$\mathscr{G}_2 Res_2(s) = \frac{h_2(x)}{s^{2\alpha+1}}\left(1 - \frac{2\alpha}{2\alpha+1}\right) - \frac{v}{s^{2\alpha}}\mathscr{G}_2\left[\partial_x^2(\varphi+\phi+\psi)^2\right]$$
$$+ \frac{w}{s^{2\alpha}}\mathscr{G}_2\left[\partial_x^4(\varphi+\phi+\psi)^2\right],$$

where

$$\varphi = \frac{2\lambda^2}{3v}\left(1 - \cosh\left(\sqrt{\frac{v}{w}}\frac{x}{2}\right)\right),$$
$$\phi = \frac{\lambda^3 t^\alpha}{3\sqrt{vw}\,\Gamma(\alpha+1)}\left(\sinh\left(\sqrt{\frac{v}{w}}\frac{x}{2}\right)\right)$$

and

$$\psi = \frac{h_2(x)t^{2\alpha}}{\Gamma(2\alpha+2)}. \tag{36}$$

After simple calculations and solving the equation $\lim_{s\to\infty} s^{2\alpha+1}\mathscr{G}_2 Res_2(x,s) = 0$ for $h_2(x)$, we have.

$$h_2(x) = -(2\alpha+1)\frac{\lambda^4}{6w}\cosh\left(\sqrt{\frac{v}{w}}\frac{x}{2}\right).$$

Repeating the previous arguments, we can obtain the coefficients of the series (25) as follows.

$$h_3(x) = \frac{(3\alpha+1)\lambda^5}{12}\sqrt{\frac{v}{w^3}}\left(\sinh\left(\sqrt{\frac{v}{w}}\frac{x}{2}\right)\right),$$
$$h_4(x) = -(4\alpha+1)\frac{\lambda^6 v}{24w^2}\cosh\left(\sqrt{\frac{v}{w}}\frac{x}{2}\right),$$
$$h_5(x) = \frac{(5\alpha+1)\lambda^7}{48}\sqrt{\frac{v^3}{w^5}}\sinh\left(\sqrt{\frac{v}{w}}\frac{x}{2}\right),$$
$$h_6(x) = -(6\alpha+1)\frac{\lambda^8 v^2}{96w^3}\cosh\left(\sqrt{\frac{v}{w}}\frac{x}{2}\right).$$

Thus, the 6th approximate solution of equation (23).

$$\mathscr{G}_2[y(x,t)] = \frac{2\lambda^2}{3vs}\left(1 - \cosh\left(\sqrt{\frac{v}{w}}\frac{x}{2}\right)\right) + \frac{(\alpha+1)\lambda^3}{3\sqrt{vw}s^{\alpha+1}}$$
$$\times \sinh\left(\sqrt{\frac{v}{w}}\frac{x}{2}\right) - \frac{(2\alpha+1)\lambda^4}{6ws^{2\alpha+1}}$$
$$\times \cosh\left(\sqrt{\frac{v}{w}}\frac{x}{2}\right) + \frac{(3\alpha+1)\lambda^5}{12s^{3\alpha+1}}\sqrt{\frac{v}{w^3}}$$
$$\times \sinh\left(\sqrt{\frac{v}{w}}\frac{x}{2}\right) - \frac{(4\alpha+1)\lambda^6 v}{24w^2 s^{4\alpha+1}}$$
$$\times \cosh\left(\sqrt{\frac{v}{w}}\frac{x}{2}\right) + \frac{(5\alpha+1)\lambda^7}{48s^{5\alpha+1}}\sqrt{\frac{v^3}{w^5}}$$
$$\times \sinh\left(\sqrt{\frac{v}{w}}\frac{x}{2}\right) - \frac{(6\alpha+1)\lambda^8 v}{96w^3 s^{6\alpha+1}}$$
$$\times \cosh\left(\sqrt{\frac{v}{w}}\frac{x}{2}\right). \tag{37}$$

Simplify expression and organize the terms in equation (37), we get.

$$\mathscr{G}_2[y(x,t)] = \frac{-2\lambda^2}{3v}\left(\cosh\left(\sqrt{\frac{v}{w}}\frac{x}{2}\right)\left[\frac{1}{s} + \frac{(2\alpha+1)\lambda^2 v^2}{2^2 w^2 s^{2\alpha+1}}\right.\right.$$
$$+ \frac{(4\alpha+1)\lambda^4 v^4}{2^4 w^4 s^{4\alpha+1}} + \frac{(6\alpha+1)\lambda^6 v^6}{2^6 w^6 s^{6\alpha+1}}\right] - \frac{1}{s}\Bigg)$$
$$+ \frac{2\lambda^2}{3v}\sinh\left(\sqrt{\frac{v}{w}}\frac{x}{2}\right)\left[\frac{(\alpha+1)\lambda}{2s^{\alpha+1}}\sqrt{\frac{v}{w}}\right.$$
$$\left.+ \frac{(3\alpha+1)\lambda^3}{2^3 s^{3\alpha+1}}\left(\sqrt{\frac{v}{w}}\right)^3 + \frac{(5\alpha+1)\lambda^5}{2^5 s^{5\alpha+1}}\left(\sqrt{\frac{v}{w}}\right)^5\right]. \tag{38}$$

After looking for equation (38), easy to notice the pattern of the terms. Hence, the series solution of equation (23) has the form.

$$\mathscr{G}_2[y(x,t)] = \frac{-2\lambda^2}{3v}\left(\cosh\left(\sqrt{\frac{v}{w}}\frac{x}{2}\right)\sum_{n=0}^{\infty}(2n\alpha+1)\frac{\left(\frac{\lambda}{2}\sqrt{\frac{v}{w}}\right)^{2n}}{s^{2n\alpha+1}}\right.$$
$$\left.- \sinh\left(\sqrt{\frac{v}{w}}\frac{x}{2}\right)\sum_{n=0}^{\infty}((2n+1)\alpha+1)\frac{\left(\frac{\lambda}{2}\sqrt{\frac{v}{w}}\right)^{2n+1}}{s^{(2n+1)\alpha+1}} - \frac{1}{s}\right) \tag{39}$$

To get the solution of the IVP (19) and (20), we apply the inverse ARA-transform of order two $\mathscr{G}_2^{-1}$ on equation (39) to get.

$$y(x,t) = \frac{-2\lambda^2}{3v}\left(\cosh\left(\sqrt{\frac{v}{w}}\frac{x}{2}\right)\sum_{n=0}^{\infty}\left(\frac{\lambda}{2}\sqrt{\frac{v}{w}}\right)^{2n}\frac{t^{2n\alpha}}{\Gamma(2n\alpha+1)}\right.$$
$$\left.- \sinh\left(\sqrt{\frac{v}{w}}\frac{x}{2}\right)\sum_{n=0}^{\infty}\left(\frac{\lambda}{2}\sqrt{\frac{v}{w}}\right)^{2n+1}\frac{t^{(2n+1)\alpha}}{\Gamma((2n+1)\alpha+1)} - 1\right). \tag{40}$$

Its worth mentioning that the solution obtained in equation (40) is identical to that introduced in references [24,61].

Table 1, shows a comparison of the ARA-RPS solution and the exact solution of example 1 with various values of $x, t$ and , $\alpha = v = w = \lambda = 1$. Also, the absolute error is presented in Table 1. In addition, the comparison of the results obtained for the exact solution corresponding to $\alpha = 1$ and the numerical solutions given by the ARA-RPSM for different values of $\alpha$, $\alpha = 0.25$, $\alpha = 0.5$, $\alpha = 0.75$ and $\alpha = 1$ are plotted in Fig. 1. Fig. 1 portray a very precise agreement of the exact solution and the ARA-RPS solutions of example 1 at different values of $\alpha$. The solution behavior in the four subfigures is the same as shown in the surface graphs in Fig. 1. There is a slight variation in decreasing or increasing the amplitudes of the solitary wave solutions due to change the values of $\alpha$.

**Example 2.** *Consider the Boussinesq equation*

$$D_t^{2\alpha}y(x,t) - y_{xx}(x,t) - \left(y^2(x,t)\right)_{xx} + \left(y(x,t)y_{xx}(x,t)\right)_{xx}$$
$$= 0, x \in \mathbb{R}, t > 0, 0 < \alpha \le 1, \tag{41}$$

*with the ICs*

$$y(x,0) = -(\gamma^2 - 1)(\cosh(x) - 1),$$
$$D_t^\alpha y(x,0) = \gamma(\gamma^2 - 1)\sinh(x). \tag{42}$$

**Solution**

Applying the ARA-transform of order two $\mathscr{G}_2$, on both sides of the equation (41).

**Table 1** The absolute errors with $v = w = \lambda = 1$ and $\alpha = 1$ for the 6th ARA-RPS solutions of the TFKG equation in example 1 with diverse values (of $x$ and $t$) and the exact solution.

| $x$ | $t$ | The exact solution | The Numerical solution | The absolute error |
|---|---|---|---|---|
| 0 | 0.25 | −0.0052155 | −0.005215 | 0 |
| 2 | 0.25 | −0.271912 | −0.271912 | 0 |
| 4 | 0.25 | −1.558058 | −1.558058 | 0 |
| 6 | 0.25 | −5.260613 | −5.260613 | 0 |
| 8 | 0.25 | −15.401151 | −15.401151 | 0 |
| 10 | 0.25 | −42.993929 | −42.993929 | 0 |
| 0 | 0.50 | −0.020942 | −0.020942 | $1.491862 \times 10^{-16}$ |
| 2 | 0.50 | −0.196456 | −0.196456 | $2.775558 \times 10^{-16}$ |
| 4 | 0.50 | −1.309459 | −1.309459 | $6.661338 \times 10^{-16}$ |
| 6 | 0.50 | −4.568853 | −4.568853 | $1.776357 \times 10^{-15}$ |
| 8 | 0.50 | −13.514867 | −13.514867 | $5.329071 \times 10^{-15}$ |
| 10 | 0.50 | −37.864312 | −37.864312 | $7.105427 \times 10^{-15}$ |
| 0 | 0.75 | −0.047427 | −0.047427 | 0 |
| 2 | 0.75 | −0.1345027 | −0.1345027 | $2.775558 \times 10^{-17}$ |
| 4 | 0.75 | −1.091777 | −1.091777 | 0 |
| 6 | 0.75 | −3.959005 | −3.959005 | $4.440892 \times 10^{-16}$ |
| 8 | 0.75 | −11.8504575 | −11.8504575 | 0 |
| 10 | 0.75 | −33.337526 | −33.337526 | $7.105427 \times 10^{-15}$ |
| 0 | 1 | −0.085084 | −0.085084 | $4.024558 \times 10^{-16}$ |
| 2 | 1 | −0.085084 | −0.085084 | $7.355228 \times 10^{-16}$ |
| 4 | 1 | −0.9016064 | −0.9016064 | $1.776357 \times 10^{-15}$ |
| 6 | 1 | −3.4215264 | −3.4215264 | $4.440892 \times 10^{-15}$ |
| 8 | 1 | −10.3818834 | −10.3818834 | $1.065814 \times 10^{-14}$ |
| 10 | 1 | −29.342747 | −29.342747 | $2.842171 \times 10^{-14}$ |

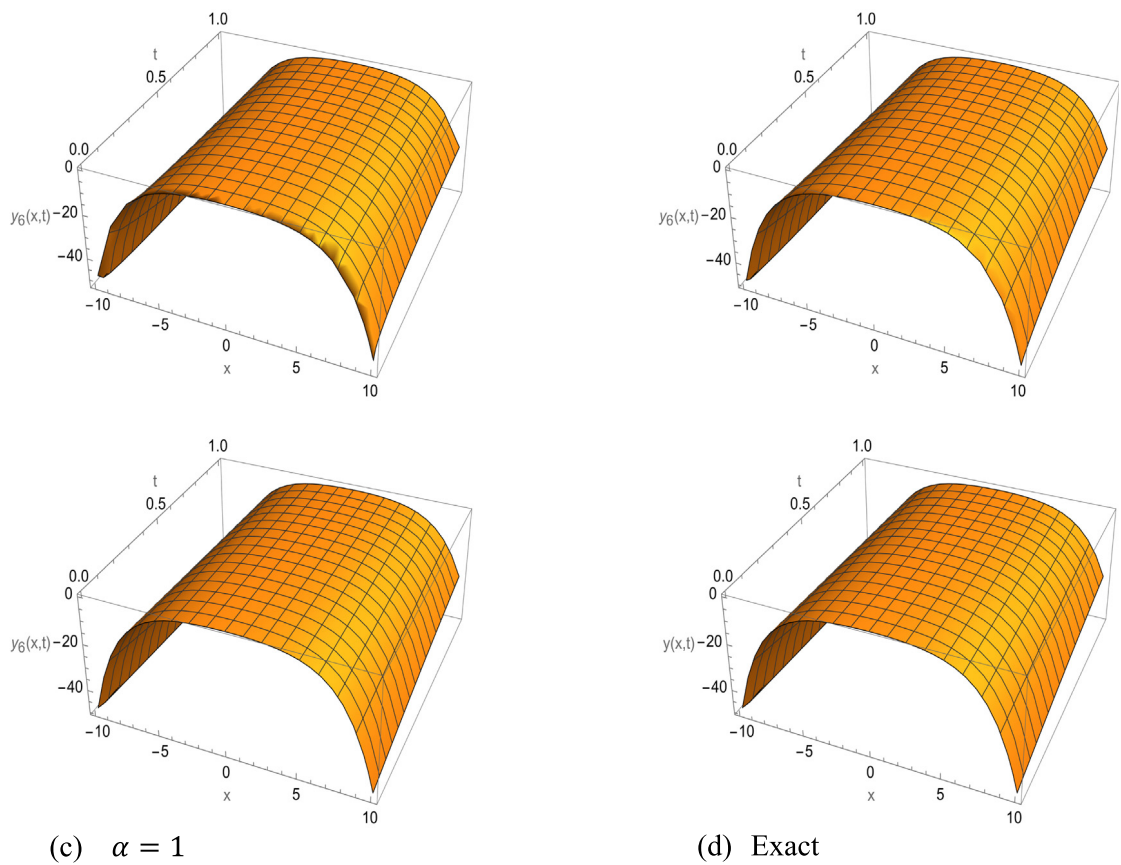

(c) $\alpha = 1$          (d) Exact

**Fig. 1** The solutions of the 6th approximation of $y(x, t)$ at different values of $\alpha$ and the exact solution in example 1.

$$\mathcal{G}_2\big[D_t^{2\alpha}y(x,t)\big] - \partial_x^2\mathcal{G}_2[y(x,t)] - \partial_x^2\mathcal{G}_2\big[y^2(x,t)\big]$$
$$+ \partial_x^2\mathcal{G}_2\big[y(x,t)\partial_x^2y(x,t)\big]$$
$$= 0. \tag{43}$$

Running the ARA-transform on equation (44), and simplifying.

$$\mathcal{G}_2[y(x,t)] - \frac{2\alpha}{s}\mathcal{G}_1[y(x,t)] + \frac{(2\alpha-1)}{s}y(x,0)$$
$$+ \frac{(\alpha-1)}{s^{\alpha+1}}D_t^\alpha y(x,0) - \frac{1}{s^{2\alpha}}\partial_x^2\mathcal{G}_2[y(x,t)]$$
$$- \frac{1}{s^{2\alpha}}\mathcal{G}_2\big[\partial_x^2\big(\mathcal{G}_1^{-1}[\mathcal{G}_2[y(x,t)]]\big)^2\big]$$
$$+ \frac{1}{s^{2\alpha}}\mathcal{G}_2\big[\partial_x^2\big(\mathcal{G}_1^{-1}[\mathcal{G}_2[(y(x,t))]]\partial_x^2\big(\mathcal{G}_1^{-1}[\mathcal{G}_2[y(x,t)]]\big)\big)\big]$$
$$= 0. \tag{44}$$

Assume that the ARA-RPS solutions of (44) have the following series representations.

$$\mathcal{G}_1[y(x,t)] = \sum_{n=0}^{\infty}\frac{h_n(x)}{(n\alpha+1)s^{n\alpha}}, \quad \mathcal{G}_2[y(x,t)] = \sum_{n=0}^{\infty}\frac{h_n(x)}{s^{n\alpha+1}}. \tag{45}$$

As the same arguments in example 1, the $k^{\text{th}}$ truncated series of the expansions in equation (45) have the form.

$$\mathcal{G}_1[y(x,t)]_k = -(\gamma^2-1)(\cosh(x)-1)$$
$$+ \frac{\gamma(\gamma^2-1)\sinh(x)}{s^\alpha} + \sum_{n=2}^{k}$$
$$\times \frac{h_n(x)}{(n\alpha+1)s^{n\alpha}}, \mathcal{G}_2[y(x,t)]_k$$
$$= \frac{-(\gamma^2-1)(\cosh(x)-1)}{s}$$
$$+ (\alpha+1)\frac{\gamma(\gamma^2-1)\sinh(x)}{s^{\alpha+1}} + \sum_{n=2}^{k}\frac{h_n(x)}{s^{n\alpha+1}}. \tag{46}$$

Define the $k^{\text{th}}$ ARA-residual function of equation (44).

$$\mathcal{G}_2Res_k(x,s) = \mathcal{G}_2[y(x,t)]_k - \frac{2\alpha}{s}\mathcal{G}_1[y(x,t)]_k$$
$$- \frac{(2\alpha-1)}{s}(\gamma^2-1)(\cosh(x)-1)$$
$$+ \frac{(\alpha-1)}{s^{\alpha+1}}\gamma(\gamma^2-1)\sinh(x)$$
$$- \frac{1}{s^{2\alpha}}\partial_x^2\mathcal{G}_2[y(x,t)]_k$$
$$- \frac{1}{s^{2\alpha}}\mathcal{G}_2\big[\partial_x^2\big(\mathcal{G}_1^{-1}[\mathcal{G}_2[y(x,t)]_k]\big)^2\big]$$
$$+ \mathcal{G}_2\big[\partial_x^2\big(\mathcal{G}_1^{-1}[\mathcal{G}_2[(y(x,t))]]\partial_x^2\big(\mathcal{G}_1^{-1}[\mathcal{G}_2[y(x,t)]_k]\big)\big)\big]. \tag{47}$$

Multiplying both sides of equation (47) by $s^{k\alpha+1}, k = 2, 3, \cdots$ and taking the limit as $s \to \infty$ to get the coefficients $h_n(x)$ in the series expansion (45) as follows.

$$h_2(x) = -(2\alpha+1)\gamma^2(\gamma^2-1)\cosh(x),$$

$$h_3(x) = (3\alpha+1)\gamma^3(\gamma^2-1)\sinh(x),$$

$$h_4(x) = -(4\alpha+1)\gamma^4(\gamma^2-1)\cosh(x),$$

$$h_5(x) = (5\alpha+1)\gamma^5(\gamma^2-1)\sinh(x),$$

$$h_6(x) = -(6\alpha+1)\gamma^6(\gamma^2-1)\cosh(x).$$

Substituting the coefficients in the series expansion of $\mathcal{G}_2[y(x,t)]$, we get the 6th ARA-approximate solution.

$$\mathcal{G}_2[y(x,t)] = -\frac{(\gamma^2-1)}{s}(\cosh(x)-1) + \frac{(\alpha+1)\gamma(\gamma^2-1)}{s^{\alpha+1}}$$
$$\times \sinh(x) - \frac{(2\alpha+1)\gamma^2(\gamma^2-1)}{s^{2\alpha+1}}\cosh(x)$$
$$+ \frac{(3\alpha+1)\gamma^3(\gamma^2-1)}{s^{3\alpha+1}}\sinh(x)$$
$$- \frac{(4\alpha+1)\gamma^4(\gamma^2-1)}{s^{4\alpha+1}}\cosh(x)$$
$$+ \frac{(5\alpha+1)\gamma^5(\gamma^2-1)}{s^{5\alpha+1}}\sinh(x)$$
$$- \frac{(6\alpha+1)\gamma^6(\gamma^2-1)}{s^{6\alpha+1}}\cosh(x). \tag{48}$$

To get the solution of the IVP (41) and (42) we apply the inverse ARA-transform of order two $\mathcal{G}_2^{-1}$ on equation (48) to get.

$$y(x,t) = -(\gamma^2-1)\left(\cosh(x)\sum_{n=0}^{\infty}\frac{\gamma^{2n}t^{2n\alpha}}{\Gamma(2n\alpha+1)} - \sinh(x)\sum_{n=0}^{\infty}\frac{\gamma^{2n+1}t^{(2n+1)\alpha}}{\Gamma((2n+1)\alpha+1)} - 1\right).$$

Which the exact solution obtained in references [24,61,63].

Table 2, shows a comparison of the ARA-RPS solution and the exact solution of example 2 with various values of $x, t$, $\alpha = 1$ and $\gamma = 2$. The absolute error is also presented in Table 2. Numerical results of example 2 with other different values of ($\gamma = 0.5$), are listed in Table 3. In addition, the comparison of the results obtained for the exact solution corresponding to $\alpha = 1$ and the numerical solutions given by the ARA-RPSM for different values of $\alpha$, $\alpha = 0.25$, $\alpha = 0.5$, $\alpha = 0.75$ and $\alpha = 1$ are plotted in Fig. 2 and Fig. 3. Fig. 2 and Fig. 3 portray a very precise agreement of the exact solution and the ARA-RPS solutions of example 2 at different values of $\alpha$.

Fig. 2 explains that by changing the fractional parameters $\alpha$, we can increase or decrease the amplitude of the solitary solutions.

**Example 3.** *Consider the time-fractional PDE*

$$D_t^{2\alpha}y(x,t) + \big(y^2(x,t)\big)_{xx} + \big(y(x,t)y_{xx}(x,t)\big)_{xx}$$
$$= 0, x \in \mathbb{R}, t > 0, 0 < \alpha \leq 1, \tag{49}$$

with the ICs

$$y(x,0) = \cosh(x) - 1, \ D_t^\alpha y(x,0) = -\sinh(x). \tag{50}$$

**Solution**

Applying the same procedure of the ARA-RPSM and using the ICs (50), the ARA-transform of order two for equation (49) can be expressed as.

$$\mathcal{G}_2[y(x,t)] - \frac{2\alpha}{s}\mathcal{G}_1[y(x,t)] + \frac{(2\alpha-1)}{s}(\cosh(x)-1)$$
$$- \frac{(\alpha-1)}{s^{\alpha+1}}\sinh(x) + \frac{1}{s^{2\alpha}}\mathcal{G}_2\big[\partial_x^2\big(\mathcal{G}_1^{-1}[\mathcal{G}_2[y(x,t)]]\big)^2\big]$$
$$+ \frac{1}{s^{2\alpha}}\mathcal{G}_2\big[\partial_x^2\big(\mathcal{G}_1^{-1}[\mathcal{G}_2[(y(x,t))]]\partial_x^2\big(\mathcal{G}_2^{-1}[\mathcal{G}_2[y(x,t)]]\big)\big)\big]$$
$$= 0. \tag{51}$$

**Table 2** The absolute errors with $\gamma = 2$ and $\alpha = 1$ for the 6th ARA-RPS solutions of the Boussinesq equation in example 2 with diverse values (of $x$ and $t$) and the exact solution.

| $x$ | $t$ | The exact solution | The Numerical solution | The absolute error |
|---|---|---|---|---|
| 0 | 0.25 | −0.382878 | −0.382878 | $2.220446 \times 10^{-16}$ |
| 2 | 0.25 | −4.057229 | −4.057229 | $1.776357 \times 10^{-15}$ |
| 4 | 0.25 | −46.718475 | −46.718474 | $1.421085 \times 10^{-14}$ |
| 6 | 0.25 | −364.044029 | −364.044029 | $1.136868 \times 10^{-13}$ |
| 8 | 0.25 | −2709.064452 | −2709.064451 | $9.094947 \times 10^{-13}$ |
| 10 | 0.25 | −20036.590357 | −20036.590357 | $1.091393 \times 10^{-11}$ |
| 0 | 0.50 | −1.629242 | −1.629242 | $2.220446 \times 10^{-16}$ |
| 2 | 0.50 | −1.629242 | −1.6292422 | $4.440892 \times 10^{-16}$ |
| 4 | 0.50 | −27.202986 | −27.202986 | $2.842171 \times 10^{-14}$ |
| 6 | 0.50 | −219.629846 | −219.629846 | $1.705303 \times 10^{-13}$ |
| 8 | 0.50 | −1641.951105 | −1641.951105 | $1.818989 \times 10^{-12}$ |
| 10 | 0.50 | −12151.626076 | −12151.626076 | $1.091394 \times 10^{-11}$ |
| 0 | 0.75 | −4.057229 | −4.057229 | $1.77636 \times 10^{-15}$ |
| 2 | 0.75 | −0.382878 | −0.382878 | $2.88658 \times 10^{-15}$ |
| 4 | 0.75 | −15.396869 | −15.396868 | $7.105427 \times 10^{-15}$ |
| 6 | 0.75 | −132.042368 | −132.042360 | $1.98952 \times 10^{-13}$ |
| 8 | 0.75 | −994.714705 | −994.714705 | $2.046363 \times 10^{-12}$ |
| 10 | 0.75 | −7369.153566 | −7369.153566 | $1.909939 \times 10^{-11}$ |
| 0 | 1 | −8.286587 | −8.286587 | $1.776357 \times 10^{-15}$ |
| 2 | 1 | 0. | $−5.329071 \times 10^{-15}$ | $5.329071 \times 10^{-15}$ |
| 4 | 1 | −8.286587 | −8.286587 | $7.105427 \times 10^{-15}$ |
| 6 | 1 | −78.924699 | −78.924699 | $3.5527137 \times 10^{-13}$ |
| 8 | 1 | −602.146908 | −602.146908 | $4.547474 \times 10^{-13}$ |
| 10 | 1 | −4468.4374838 | −4468.437484 | $1.4551915 \times 10^{-11}$ |

**Table 3** The absolute errors with $\gamma = 0.5$ and $\alpha = 1$ for the 6th ARA-RPS solutions of the Boussinesq equation in example 2 with diverse values (of $x$ and $t$) and the exact solution.

| $x$ | $t$ | The exact solution | The Numerical solution | The absolute error |
|---|---|---|---|---|
| 0 | 0.25 | 0.005867 | 0.005867 | $2.081668 \times 10^{-17}$ |
| 2 | 0.25 | 1.752815 | 1.752815 | $2.220446 \times 10^{-16}$ |
| 4 | 0.25 | 17.326295 | 17.326295 | 0. |
| 6 | 0.25 | 132.760301 | 132.760301 | $2.842171 \times 10^{-14}$ |
| 8 | 0.25 | 985.757464 | 985.757464 | 0. |
| 10 | 0.25 | 7288.607959 | 7288.607959 | $9.094947 \times 10^{-13}$ |
| 0 | 0.50 | 0.02356 | 0.02356 | $1.769418 \times 10^{-16}$ |
| 2 | 0.50 | 1.473141 | 1.473141 | $6.661338 \times 10^{-16}$ |
| 4 | 0.50 | 15.204225 | 15.204225 | $5.329071 \times 10^{-15}$ |
| 6 | 0.50 | 117.072691 | 117.072691 | $7.105427 \times 10^{-14}$ |
| 8 | 0.50 | 869.839817 | 869.839817 | $2.273737 \times 10^{-13}$ |
| 10 | 0.50 | 6432.085825 | 6432.085825 | $1.818989 \times 10^{-12}$ |
| 0 | 0.75 | 0.053355 | 0.053355 | $2.775558 \times 10^{-17}$ |
| 2 | 0.75 | 1.228249 | 1.228249 | 0 |
| 4 | 0.75 | 13.3317646 | 13.331765 | $3.552714 \times 10^{-15}$ |
| 6 | 0.75 | 103.228459 | 103.228459 | 0. |
| 8 | 0.75 | 767.542857 | 767.542857 | $2.273737 \times 10^{-13}$ |
| 10 | 0.75 | 5676.207696 | 5676.207696 | 0 |
| 0 | 1 | 0.095719 | 0.095719 | $5.551115 \times 10^{-17}$ |
| 2 | 1 | 1.014307 | 1.014307 | $4.440892 \times 10^{-16}$ |
| 4 | 1 | 11.679619 | 11.679619 | $3.552714 \times 10^{-15}$ |
| 6 | 1 | 91.011007 | 91.011007 | $2.842171 \times 10^{-14}$ |
| 8 | 1 | 677.266113 | 677.266113 | $2.273737 \times 10^{-13}$ |
| 10 | 1 | 5009.147589 | 5009.147589 | $2.728484 \times 10^{-12}$ |

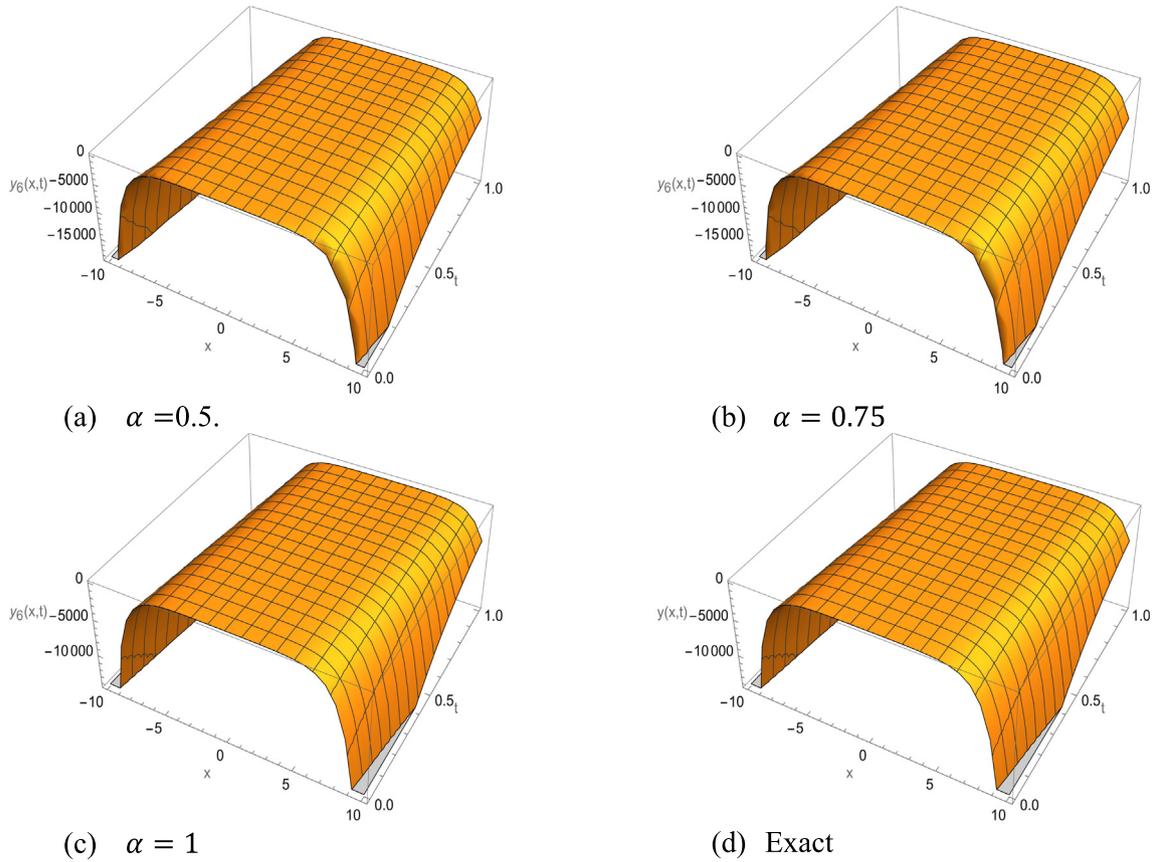

(a)   $\alpha = 0.5$.

(b)   $\alpha = 0.75$

(c)   $\alpha = 1$

(d)   Exact

**Fig. 2**   The solutions of the 6th approximation of $y(x,t)$ at different values of $\alpha$ and the exact solution in example 2.

Assume that the solution of equation (59) has the following series representations.

$$\mathscr{G}_1[y(x,t)] = \sum_{n=0}^{\infty} \frac{h_n(x)}{(n\alpha+1)s^{n\alpha}}$$

$$\mathscr{G}_2[y(x,t)] = \sum_{n=0}^{\infty} \frac{h_n(x)}{s^{n\alpha+1}}, x \in \mathbb{R}, s > 0. \qquad (52)$$

Using the ICs (50), the $k^{\text{th}}$ truncated series of equation (52).

$$\mathscr{G}_2[y(x,t)]_k = \frac{\cosh(x)-1}{s} - \frac{(\alpha+1)\sinh(x)}{s^{\alpha+1}} + \sum_{n=2}^{k} \frac{h_n(x)}{s^{n\alpha+1}}.$$

We define the $k^{\text{th}}$ ARA-residual function of equation (51) as.

$$\begin{aligned}
\mathscr{G}_2 Res_k(x,s) &= \mathscr{G}_2[y(x,t)]_k - \frac{2\alpha}{s}\mathscr{G}_1[y(x,t)]_k \\
&\quad + \frac{(2\alpha-1)}{s}(\cosh(x)-1) - \frac{(\alpha-1)}{s^{\alpha+1}}\sinh(x) \\
&\quad + \frac{1}{s^{2\alpha}}\mathscr{G}_2\left[\partial_x^2\left(\mathscr{G}_2^{-1}\left[\mathscr{G}_2[y(x,t)]_k\right]\right)^2\right] \\
&\quad - \mathscr{G}_2\left[\partial_x^2\left(\mathscr{G}_2^{-1}\left[\mathscr{G}_2[(y(x,t))]\right]\right)\partial_x^2\left(\mathscr{G}_2^{-1}\left[\mathscr{G}_2[y(x,t)]_k\right]\right)\right].
\end{aligned} \qquad (53)$$

In a similar way to the previous examples, we solve the equations.

$$\lim_{s\to\infty} s^{k\alpha+1}\mathscr{G}_2 Res_k(x,s) = 0, k = 2, 3, \cdots.$$

The coefficients $h_n(s)$ of the series expansion (52) will be.

$$h_2(x) = (2\alpha+1)\cosh(x),$$

$$h_3(x) = -(3\alpha+1)\sinh(x),$$

$$h_4(x) = (4\alpha+1)\cosh(x),$$

$$h_5(x) = -(5\alpha+1)\sinh(x),$$

$$h_6(x) = (6\alpha+1)\cosh(x).$$

Thus, we have the following series solution of equation (51) in the form.

$$\begin{aligned}
\mathscr{G}_2[y(x,t)] &= \frac{\cosh(x)-1}{s} - \frac{(\alpha+1)\sinh(x)}{s^{\alpha+1}} \\
&\quad + \frac{(2\alpha+1)\cosh(x)}{s^{2\alpha+1}} - \frac{(3\alpha+1)\sinh(x)}{s^{3\alpha+1}} \\
&\quad + \frac{(4\alpha+1)\cosh(x)}{s^{4\alpha+1}} - \frac{(5\alpha+1)\sinh(x)}{s^{5\alpha+1}} \\
&\quad + \frac{(6\alpha+1)\cosh(x)}{s^{6\alpha+1}}.
\end{aligned} \qquad (54)$$

Running the inverse ARA-transform of order two $\mathscr{G}_2^{-1}$ on equation (54), we get the solution of the IVP (49) and (50) in the original space as.

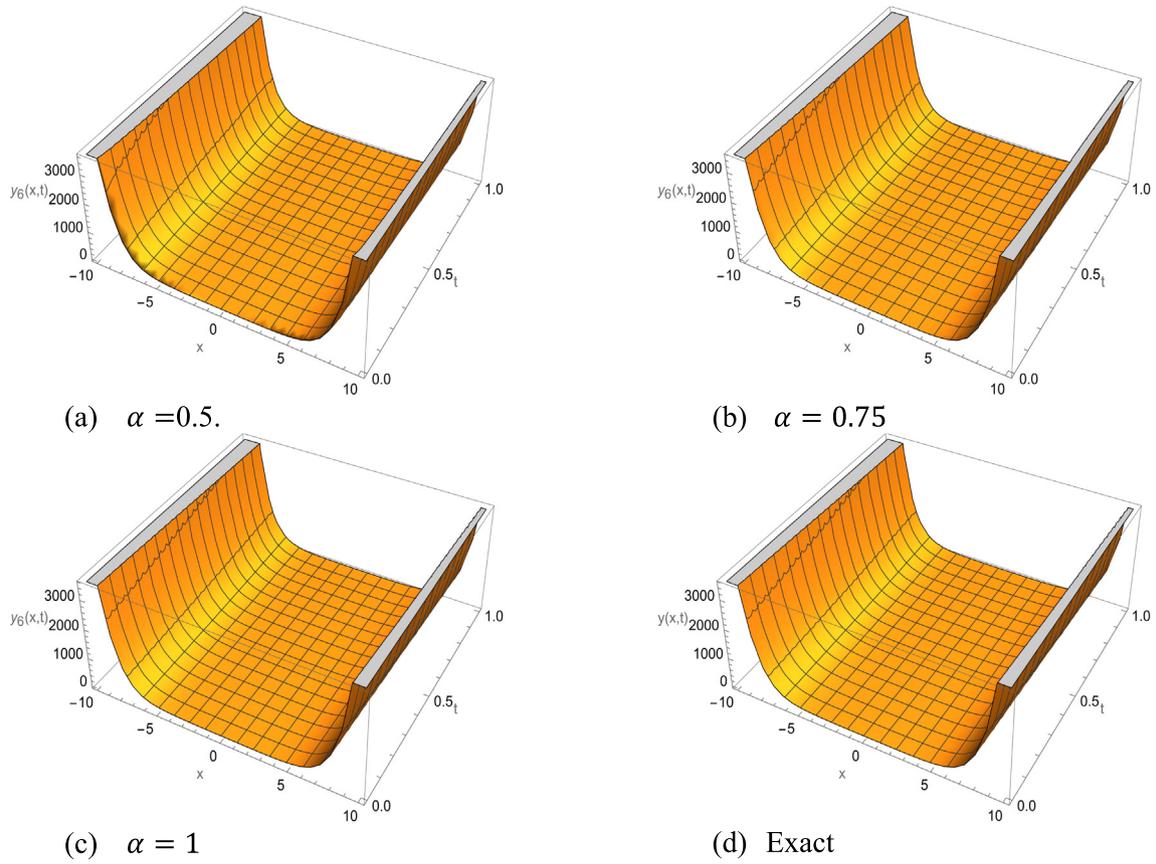

**(a)** $\alpha = 0.5$.

**(b)** $\alpha = 0.75$

**(c)** $\alpha = 1$

**(d)** Exact

**Fig. 3** The solutions of the 6th approximation of $y_6(x,t)$ at different values of $\alpha$ and the exact solution in example 2.

$$y(x,t) = \cosh(x) \sum_{n=0}^{\infty} \frac{t^{2n\alpha}}{\Gamma(2n\alpha+1)} - \sinh(x) \sum_{n=0}^{\infty} \times \frac{t^{(2n+1)\alpha}}{\Gamma((2n+1)\alpha+1)} - 1.$$

Putting $\alpha = 1$, we get the exact solitary solution of the IVP (49) and (50) as.

$$y(x,t) = 2\sinh^2\left(\frac{1}{2}(x-t)\right).$$

Which is the same solution obtained in reference [24,51,53]. Table 4, shows a comparison of the ARA-RPS solution and the exact solution of example 3 with various values of $x$, $t$ and $\alpha = 1$. Also, the absolute error is presented in Table 4. In addition, the comparison of the results obtained for the exact solution corresponding to $\alpha = 1$ and the numerical solutions given by the ARA-RPSM for different values of $\alpha$, $\alpha = 0.25$, $\alpha = 0.5$, $\alpha = 0.75$ and $\alpha = 1$ are plotted in Fig. 4. Fig. 4 portrays a very precise agreement of the exact solution and the ARA-RPS solutions of example 3 at different values of $\alpha$.

**Example 4.** Consider the time-fractional PDE

$$D_t^{\alpha} y(x,t) - \left(y^3(x,t)\right)_x + \left(y^3(x,t)\right)_{xxx} = 0, x \in \mathbb{R}, t > 0,$$
$$0 < \alpha \leq 1, \tag{55}$$

*with the IC*

$$y(x,0) = \sqrt{\frac{3}{2}} \sinh\left(\frac{x}{3}\right). \tag{56}$$

**Solution**

In a similar argument in the previous examples.

We applying the ARA transform of order two $\mathscr{G}_2$ of equation (55) and use the IC (56), to get.

$$\mathscr{G}_2[y(x,t)] - \frac{\alpha}{s}\mathscr{G}_1[y(x,t)] + \frac{(\alpha-1)}{s}\left(\sqrt{\frac{3}{2}}\sinh\left(\frac{x}{3}\right)\right)$$
$$- \frac{1}{s^{\alpha}}\mathscr{G}_2\left[\partial_x\left(\mathscr{G}_2^{-1}[\mathscr{G}_2[y(x,t)]]\right)^3\right]$$
$$+ \frac{1}{s^{\alpha}}\mathscr{G}_2\left[\partial_x^3\left(\mathscr{G}_2^{-1}[\mathscr{G}_2[(y(x,t))]]\right)^3\right]$$
$$= 0. \tag{57}$$

Define the $k^{\text{th}}$ truncated series solutions of equation (57).

$$\mathscr{G}_1[y(x,t)]_k = \sqrt{\frac{3}{2}}\sinh\left(\frac{x}{3}\right) + \sum_{n=1}^{k}\frac{h_n(x)}{(n\alpha+1)s^{n\alpha}},$$

$$\mathscr{G}_2[y(x,t)]_k = \sqrt{\frac{3}{2}}\frac{\sinh\left(\frac{x}{3}\right)}{s} + \sum_{n=1}^{k}\frac{h_n(x)}{s^{n\alpha+1}}, x \in \mathbb{R}, s > 0. \tag{58}$$

and the $k^{\text{th}}$ ARA-residual function of equation (57) can be written as.

**Table 4**  The absolute errors with $\alpha = 1$ for the 6th ARA-RPS solutions of the TFPDE in example 3 with diverse values (of $x$ and $t$) and the exact solution.

| $x$ | $t$ | The exact solution | The Numerical solution | The absolute error |
|-----|-----|-----|-----|-----|
| 0 | 0.25 | 0.031413 | 0.031413 | $1.387779 \times 10^{-17}$ |
| 2 | 0.25 | 1.964188 | 1.964188 | $2.220446 \times 10^{-16}$ |
| 4 | 0.25 | 20.2723 | 20.2723 | 0 |
| 6 | 0.25 | 156.096922 | 156.096922 | $2.842171 \times 10^{-14}$ |
| 8 | 0.25 | 1159.786423 | 1159.786423 | $2.273737 \times 10^{-13}$ |
| 10 | 0.25 | 8576.114434 | 8576.114434 | $1.818989 \times 10^{-12}$ |
| 0 | 0.50 | 0.127626 | 0.127626 | $7.494005 \times 10^{-16}$ |
| 2 | 0.50 | 1.35241 | 1.35241 | $2.88658 \times 10^{-15}$ |
| 4 | 0.50 | 15.572825 | 15.572825 | $2.131628 \times 10^{-14}$ |
| 6 | 0.50 | 121.34801 | 121.34801 | $1.563194 \times 10^{-13}$ |
| 8 | 0.50 | 903.021484 | 903.021484 | $1.136868 \times 10^{-12}$ |
| 10 | 0.50 | 6678.863452 | 6678.863452 | $1.000444 \times 10^{-11}$ |
| 0 | 0.75 | 0.294683 | 0.294683 | $2.048361 \times 10^{-13}$ |
| 2 | 0.75 | 0.888424 | 0.888424 | $7.329692 \times 10^{-13}$ |
| 4 | 0.75 | 11.914557 | 11.914557 | $5.307754 \times 10^{-12}$ |
| 6 | 0.75 | 94.285758 | 94.285758 | $3.923617 \times 10^{-11}$ |
| 8 | 0.75 | 703.052779 | 703.052779 | $2.899014 \times 10^{-10}$ |
| 10 | 0.75 | 5201.282906 | 5201.282906 | $2.14277 \times 10^{-9}$ |
| 0 | 1 | 0.543081 | 0.543081 | $1.151879 \times 10^{-11}$ |
| 2 | 1 | 0.543081 | 0.543081 | $4.055167 \times 10^{-11}$ |
| 4 | 1 | 9.067662 | 9.067662 | $2.936122 \times 10^{-10}$ |
| 6 | 1 | 73.209949 | 73.209949 | $2.168719 \times 10^{-9}$ |
| 8 | 1 | 547.317035 | 547.317035 | $1.602473 \times 10^{-8}$ |
| 10 | 1 | 4050.542025 | 4050.542025 | $1.184035 \times 10^{-7}$ |

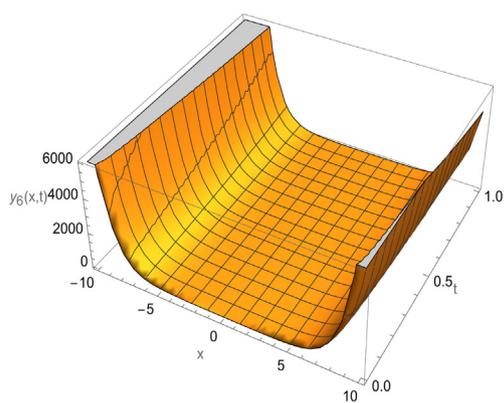

(a)  $\alpha = 0.5$.

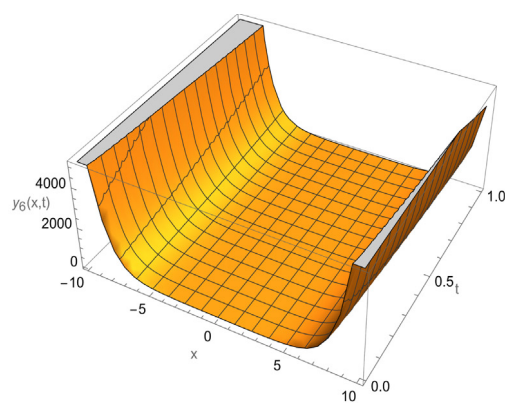

(b)  $\alpha = 0.75$

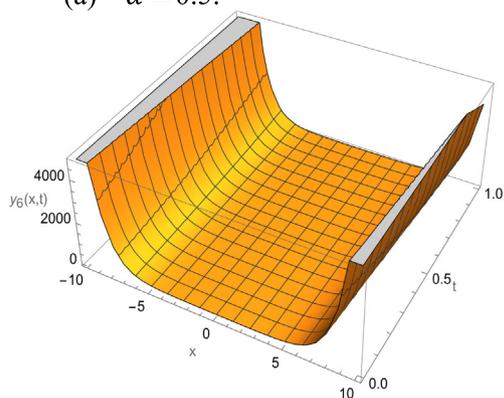

(c)  $\alpha = 1$

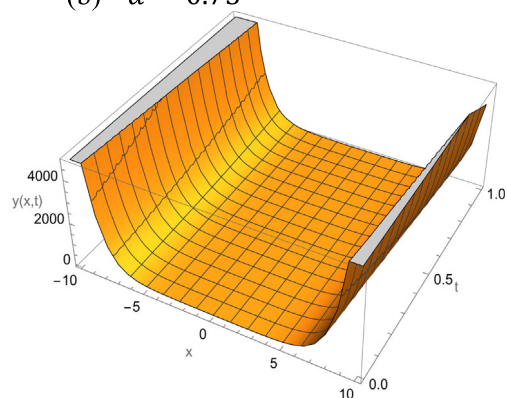

(d)  Exact

**Fig. 4**  The solutions of the 6th approximation of $y(x, t)$ at different values of $\alpha$ and the exact solution in example 3.

**Table 5** The absolute errors with $\alpha = 1$ for the 6th ARA-RPS solutions of the TFPDE in example 4 with diverse values (of $x$ and $t$) and the exact solution.

| $x$ | $t$ | The exact solution | The Numerical solution | The absolute error |
|---|---|---|---|---|
| 0 | 0.25 | −0.102180 | −0.102180 | $6.782491 \times 10^{-12}$ |
| 2 | 0.25 | 0.755647 | 0.755647 | $8.295808 \times 10^{-12}$ |
| 4 | 0.25 | 1.961942 | 1.961942 | $1.363465 \times 10^{-11}$ |
| 6 | 0.25 | 4.072989 | 4.072989 | $2.526157 \times 10^{-11}$ |
| 8 | 0.25 | 8.0623 | 8.0623 | $4.853895 \times 10^{-11}$ |
| 10 | 0.25 | 15.769549 | 15.769549 | $9.419843 \times 10^{-11}$ |
| 0 | 0.50 | −0.20507 | −0.20507 | $8.684082 \times 10^{-10}$ |
| 2 | 0.50 | 0.638209 | 0.638209 | $1.055668 \times 10^{-9}$ |
| 4 | 0.50 | 1.7758 | 1.7758 | $1.729751 \times 10^{-9}$ |
| 6 | 0.50 | 3.732301 | 3.732301 | $3.201509 \times 10^{-9}$ |
| 8 | 0.50 | 7.409957 | 7.409957 | $6.149647 \times 10^{-9}$ |
| 10 | 0.50 | 14.504724 | 14.504724 | $1.193371 \times 10^{-8}$ |
| 0 | 0.75 | −0.309386 | −0.309386 | $1.484473 \times 10^{-8}$ |
| 2 | 0.75 | 0.525205 | 0.525205 | $1.793493 \times 10^{-8}$ |
| 4 | 0.75 | 1.601995 | 1.601995 | $2.929584 \times 10^{-8}$ |
| 6 | 0.75 | 3.417546 | 3.417546 | $5.416657 \times 10^{-8}$ |
| 8 | 0.75 | 6.809102 | 6.809102 | $1.040163 \times 10^{-7}$ |
| 10 | 0.75 | 13.340684 | 13.340684 | $2.018332 \times 10^{-7}$ |
| 0 | 1 | −0.415851 | −0.415851 | $1.112851 \times 10^{-7}$ |
| 2 | 1 | 0.415851 | 0.415851 | $1.336203 \times 10^{-7}$ |
| 4 | 1 | 1.439322 | 1.439322 | $2.175747 \times 10^{-7}$ |
| 6 | 1 | 3.126538 | 3.126538 | $4.018639 \times 10^{-7}$ |
| 8 | 1 | 6.25556 | 6.25556 | $7.714731 \times 10^{-7}$ |
| 10 | 1 | 12.269341 | 12.269341 | $1.496848 \times 10^{-6}$ |

$$\mathscr{G}_2 Res_k(x,s) = \mathscr{G}_2[y(x,t)]_k - \frac{\alpha}{s}\mathscr{G}_1[y(x,t)]_k + \sqrt{\frac{3}{2}}\frac{\alpha - 1}{s}$$
$$\times \sinh\left(\frac{x}{3}\right)$$
$$- \frac{1}{s^\alpha}\mathscr{G}_2\left[\partial_x\left(\mathscr{G}_2^{-1}\left[\mathscr{G}_2[y(x,t)]_k\right]\right)^3\right]$$
$$+ \frac{1}{s^\alpha}\mathscr{G}_2\left[\partial_x^3\left(\mathscr{G}_2^{-1}\left[\mathscr{G}_2[(y(x,t))]_k\right]\right)^3\right]. \quad (59)$$

We solve the following equations.

$$\lim_{s\to\infty} s^{k\alpha+1}\mathscr{G}_2 Res_k(x,s) = 0, k = 1, 2, 3, \cdots.$$

Performing the above steps, we can express the series solution of equation (55) in the ARA space as follows.

$$\mathscr{G}_2[y(x,t)] = \sqrt{\frac{3}{2}}$$
$$\times \sinh\left(\frac{x}{3}\right)\left(\frac{1}{s} + \frac{2\alpha + 1}{3^2 s^{2\alpha+1}} + \frac{4\alpha + 1}{3^4 s^{4\alpha+1}} + \frac{6\alpha + 1}{3^6 s^{6\alpha+1}}\right)$$
$$- \sqrt{\frac{3}{2}}$$
$$\times \cosh\left(\frac{x}{3}\right)\left(\frac{\alpha + 1}{3 s^{\alpha+1}} + \frac{3\alpha + 1}{3^3 s^{3\alpha+1}} + \frac{5\alpha + 1}{3^5 s^{5\alpha+1}}\right). \quad (60)$$

Hence, the solitary solution of the IVP (55) and (56) in the original space can be obtained by applying $\mathscr{G}_2^{-1}$ on equation (60) to get.

$$y(x,t) = \sqrt{\frac{3}{2}}\sinh\left(\frac{x}{3}\right)\cosh\left(\frac{t^\alpha}{3}\right)$$
$$- \sqrt{\frac{3}{2}}\cosh\left(\frac{x}{3}\right)\sinh\left(\frac{t^\alpha}{3}\right). \quad (61)$$

It's good to mention here that the obtained solution in equation (61) is identical to that obtained in references [64].

The exact solution when $\alpha = 1$, becomes.

$$y(x,t) = \sqrt{\frac{3}{2}}\sinh\left(\frac{x-t}{3}\right).$$

Table 5, shows a comparison of the ARA-RPS solution and the exact solution of example 4 with various values of $x$, $t$ and $\alpha = 1$. Also, the absolute error is presented in Table 5. In addition, the comparison of the results obtained for the exact solution corresponding to $\alpha = 1$ and the numerical solutions given by the ARA-RPSM for different values of $\alpha$, $\alpha = 0.25$, $\alpha = 0.5$, $\alpha = 0.75$ and $\alpha = 1$ are plotted in Fig. 5. Fig. 5 portrays a very precise agreement of the exact solution and the ARA-RPS solutions of example 4 at different values of $\alpha$.

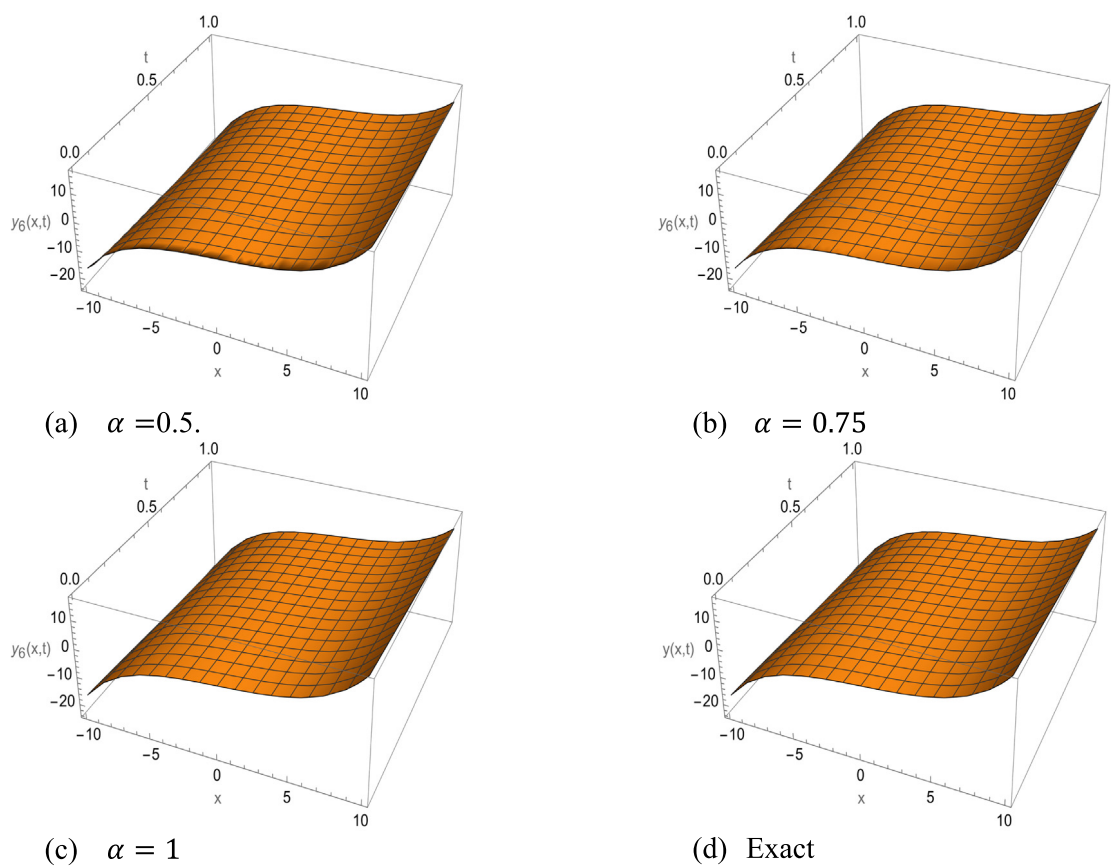

**Fig. 5** The solutions of the 6th approximation of $y(x, t)$ at different values of α and the exact solution in example 4.

## 5. Conclusion

In this research, we presented a new method called the ARA-RPS. ARA-residual power series method which is implemented to construct approximate solitary series solutions to nonlinear PDEs of time fractional order. We introduced the technique of the proposed method through an algorithm that guide the authors to use it. We presented four numerical experiments and sketch the solutions with different values of α, to show applicability and simplicity of the ARA-RPSM. We compared our results with that obtained by other methods and the exact solution. As a future work, we plane to solve linear and nonlinear fractional integral equations by the ARA-RPSM.

### Declaration of Competing Interest

The authors declare that they have no known competing financial interests or personal relationships that could have appeared to influence the work reported in this paper.

### Acknowledgments

The authors express their gratitude to the dear unknown referees and the editor for their helpful suggestions, which improved the final version of this paper.

This research is funded by the Deanship of Research in Zarqa University, Jordan